# INTEGRAL – a status report

**Christoph Winkler[1]**

*Research and Science Support Department, Astrophysics and Fundamental Missions Division*
*ESA-ESTEC, 2201 AZ Noordwijk, The Netherlands*
*E-mail:* `cwinkler@rssd.esa.int`

The ESA gamma-ray observatory INTEGRAL, launched on 17 October 2002, continues to produce a wealth of discoveries and new results on compact high energy Galactic objects, nuclear gamma-ray line emission, diffuse line and continuum emission, cosmic background radiation, AGN, high energy transients and sky surveys. The observing programme, fully open to the scientific community at large, is built from the community's feedback to the Announcements of Opportunity, issued about once per year. The mission's technical status is healthy and INTEGRAL is continuing its scientific operations well beyond its 5-year technical design lifetime. This paper will briefly summarize the overall current status.

---

[1] Speaker

# 1. The INTEGRAL mission – an overview

The ESA observatory INTEGRAL [1] is dedicated to the fine spectroscopy (2.5 keV FWHM @ 1 MeV) and fine imaging (angular resolution: 12 arcmin FWHM) of celestial gamma-ray sources in the energy range 15 keV to 10 MeV with concurrent source monitoring in the X-ray (3-35 keV) and optical (V-band, 550 nm) bands. INTEGRAL, with a total launch mass of about 4 tons was launched by a four-stage PROTON from Baikonur/Kazakhstan on 17 October 2002. The orbit is characterized by a high perigee in order to provide long periods of uninterrupted observations with nearly constant background and away from trapped radiation (electron and proton radiation belts). The orbital parameters at the beginning of the mission were: 72-hour orbit with an inclination of 52.2 degrees, a height of perigee of 9,050 km and a height of apogee of 153,657 km.

Owing to background radiation effects in the high-energy detectors, scientific observations are carried out while the satellite is above a nominal altitude of typically 40,000 to 60,000 km. This means that most of the time spent in the orbit provided by the PROTON can be used for scientific observations, about 210 ksec per revolution. An on-board particle radiation monitor allows the continuous assessment of the radiation environment local to the spacecraft. INTEGRAL carries two main gamma-ray instruments, the spectrometer SPI [2], optimized for the high-resolution (2.5 keV FWHM @ 1 MeV) gamma-ray line spectroscopy (20 keV - 8 MeV), and the imager IBIS [3], optimized for high-angular resolution (12 arcmin FWHM) imaging (15 keV - 10 MeV). Two monitors, JEM-X [4] in the (3-35) keV X-ray band, and OMC [5] in the optical Johnson V-band complement the payload. All instruments are co-aligned with overlapping fully coded field--of--views ranging from 4.8º diameter (JEM-X), 5ºx5º (OMC), to 9ºx9º (IBIS) and 16º corner-to-corner (SPI), and they are operated simultaneously, providing the observer with data from all 4 instruments.

The Mission Operations Centre at ESOC (Darmstadt/Germany), using ground stations in Redu (Belgium) and Goldstone (USA) performs all standard spacecraft and payload operations and maintenance tasks. The INTEGRAL Science Operations Centre in Madrid (Spain) is responsible for the science operations planning including the implementation of Target of Opportunity observations within the pre-planned observing programme. The INTEGRAL Science Data Centre in Versoix (Switzerland) receives the science telemetry for near-real time monitoring, standard science analysis and archive ingestion.

# 2. The INTEGRAL Observatory

INTEGRAL was conceived from its initial study phase in 1989 as an observatory-type mission with a nominal lifetime of 2 years, and a technical life time of 5 years. The observing programme was initially shared between an open-time "General Programme" with observations awarded to PIs following a standard AO-peer review process and a guaranteed-time "Core

Programme". The share of observing time between Core and General Programmes varied from 35%/65% after launch to a fully open (100%) General Programme in 2008, at the end of the 6th year in operation. Typical observations last from few 10's ksec up to about one or few Msec. In 2006 (for the cycle of AO-4 observations), ESA introduced a new "open-time" programme element, the so-called "Key Programme" (KP) which is characterized by a scientific investigation requiring a significant amount of annual observing time (one to few Ms) in order to fulfill INTEGRAL's prime scientific objectives.

Thanks to the large FOV, it is therefore possible that many scientists can benefit through participation in the KP observation, by analysing scientific data from specific point sources, or extended areas (only), which are contained in the KP sky area. Therefore, scientists interested in different sources or various astrophysical phenomena will be able to benefit from the same observation, and through this, maximize the mission's scientific return. This can technically be accomplished by submitting so-called data rights proposals, which give – to the successful proposer – the exclusive data rights on individual point sources, or sky areas, but not (by default) automatically for the entire dataset of the observation which contains these (and more) sources in its FOV. With the AO-7 cycle of observations, starting in Fall 2009, the data rights proposals can also apply to participate in all other non-KP observations, with the exception of ToO observations, generalizing thereby the Key Programme concept. A single Time Allocation Committee evaluates all proposals for open-time General Programme (including ToOs) and Key Programme observations, as well as data rights proposals to be associated with approved observations, on their scientific merit. The first operational years were dominated by observations of the Galactic bulge and plane regions, largely also thanks to the high priority Core Programme observations. During the recent years, the sky exposure has started to open up for higher galactic latitudes. This, for example, has led to a significant increase in the number of detected extragalactic sources as listed in the 4th IBIS/ISGRI survey [6] compared to earlier IBIS survey catalogues. As of October 2009, INTEGRAL has spent about 168 Ms on science observations including dozens of Target of Opportunity observations (totaling about 2 Ms per year), which provide highly important serendipitous science.

### 3. Technical Status

The INTEGRAL spacecraft continues to operate flawlessly, seven years after launch. INTEGRAL is in its extended mission phase, beyond its 5 year technical lifetime. The amount of on-board fuel, which is used for attitude control, orbit maintenance is about 130 kg, with an average consumption of about 100 g/week. Likewise, the solar array power margin is very healthy. The spacecraft is still being operated in its nominal configuration which, thanks to it's solar aspect angle constraint of ±40°, allows a good sky visibility at any point in time.

Solar and lunar gravitation are influencing the orbital parameters. For example, the perigee height evolved from 9000 km at the start of the mission to 13000 km in 2007, to 7300 km in October 2009, and will reach a minimum of 3000 km in 2012, after which it will increase again to a local maximum of 10000 km in 2016 with subsequent decrease. During the approach to the 2012 minimum, the background rates will be monitored closely. A so-called Δ-v manoeuvre could raise the spacecraft into an orbit with a minimum perigee of 4000 km (reached one year later, in 2013), using about 50 kg of on-board fuel. Background data from the SIGMA/Granat mission, which was in a quite similar orbital radiation environment, however, indicate that it may not be necessary to perform a Δ-v manoeuvre at all [7].

The payload is in a very stable configuration. However, unfortunately, the SPI instrument had "lost" it's 3rd Ge detector (out of 19 in total) in February 2009. The cause for the loss is not understood, possibly linked to the pre-amplifier electronic circuit. Two other Ge detectors had to be switched off during the early mission phase (2003, 2004). However, it should be emphasized that SPI is fully operational in space for more than 7 years using 16 out of 19 Ge detectors, with 92% of its pre-launch sensitivity.

### 4. Science community interfaces

How can the science community interface with the INTEGRAL mission ?

(1) INTEGRAL is operated as an observatory. The observing programme, 100% open to the community at large, is built from observing proposals which have been peer reviewed by the Time Allocation Committee (TAC). Submission of proposals is invited following an Announcement of Opportunity (AO), issued by ESA about once a year. The community's feedback to INTEGRAL AO's shows that INTEGRAL observing programmes are oversubscribed in time by about a factor 6 (on average, during the extended mission phase). Observations have resulted in more than 1100 publications since launch (until August 2009) out of which about 500 appeared in refereed journals.

(2) Scientists can access public and private data from various archives : (i) The INTEGRAL science data centre (ISDC) is providing consolidated and near-real time data. (ii) Archival data are available at ISDC, and at the INTEGRAL science operations centre (ISOC), at the INTEGRAL Guest Observer Facility at NASA/GSFC, and at the Russian Science Data Centre at IKI (Moscow). (iii) Instantaneous public data are available e.g. from the SPI anti-coincidence susbsytem (off-axis GRB and SGR light-curves), data from Galactic Bulge monitoring programs, data from GRB detected in the FOV (e.g. light-curve, fluence), data from in-flight calibration observations etc.

(3) Target of Opportunity observations which provide important serendipitous science are possible via the execution of TAC approved TOO proposals, which will be scheduled, once their specific scientific and operational criteria have been fulfilled. However, in case of a new (unexpected) event, scientists can always alert the INTEGRAL science operations team via a TOO alert1 in order to consider an observation which is timely and of scientific relevance, even if it is not included in the database of accepted observing programmes of the on-going AO cycle.

(4) The scientific users community at large can interface with the INTEGRAL project via two important committees[2]. Scientists who represent the users community at large are members in both committees. Membership is usually on a two-year rotational basis. (a) The Time Allocation Committee in charge of the peer review of all observing and data rights proposals. All submitted proposals will be reviewed on their scientific merit. Successful proposals will be recommended to ESA for implementation. (b) The INTEGRAL Users Group, which was established by ESA in 2005 as a merger with the original INTEGRAL Science Working Team. Its prime objectives are, to: (i) Maximize the scientific return of INTEGRAL; (ii) Ensure that the INTEGRAL observatory is satisfying the objectives of the scientific community at large, and (iii) Act as a focus for the interests of the scientific community in INTEGRAL and act as an advocate for INTEGRAL within that community.

## 5. Scientific achievements

A selection of key scientific results obtained by INTEGRAL since launch is briefly summarized below, in <u>chronological</u> order[3]:

• The serendipitous detection of the INTEGRAL source IGR J16318-4848 has led to a new class of HMXB, namely obscured super-giant systems, where the accreting compact object is embedded in the very dense (with typical $N_H \sim 10^{23..24}$ cm$^{-2}$) wind of the super-giant companion. These sources have been observed predominantly towards the Galactic bulge and and Norma, Scutum spiral arms [8], [9], [10].

• The diffuse continuum emission, 20-200 keV, from the galactic central radian could be resolved into integrated emission from discrete point sources, leaving only a small percentage for the truly diffuse emission component [11].

• The first all-sky map in the light of the 511 keV γ-ray line, produced by the annihilation of electrons and positrons presents a unique, and yet unexplained, view of our Galaxy. The annihilation rate in the bulge is: $1.5 \times 10^{43}$s$^{-1}$. The high Bulge/Disk luminosity ratio (3…9) imposes severe constraints on the principal positron source [12].

• The 511 keV line itself is un-shifted and the excellent SPI high-resolution data provide most stringent constraints on the temperature and ionization state of the ISM: the line consists of a narrow (1.3 keV FWHM) component resulting from annihilation in a warm, ionized ISM, and a broader (5.3 keV FWH) component resulting from in-flight annihilation in a warm, neutral ISM. The Positronium fraction is 97% [30, 31].

• Radioactivity from the nucleosynthesis in massive stars is manifested in the nuclear line emission at 1.8 MeV from 26Al. The line centroids vary with Galactic longitude as expected from differential rotation in the Galaxy thus reflecting the current massive-star population throughout the entire Galaxy. These observations allow to independently determine the current

---

[2] *See for more information: http://www.sciops.esa.int/integral*
[3] *http://www.sciops.esa.int/index.php?project=INTEGRAL&page=Publications*

Galactic core-collapse SN rate: (1.9 ± 1.1) per 100 y, and to derive a star formation rate of (4 ± 2) $M_{solar}$/y or about 7.5 stars per year [13].

• Serendipitous observations of anomalous X-ray pulsars (AXP) show an unexpected very hard power-law spectrum unveiling previously unknown early phases of magnetars [14].

• Thanks to the unique combination of INTEGRAL's very large FOV, good sensitivity, broad energy range and long exposures, a new sub-class of super-giant HMXB, the super-giant fast X-ray transients was discovered, characterized by short, bright outbursts on short timescales (~ hr). Many more HMXB of this type are expected to be hidden throughout our Galaxy [15].

• A unique observation of the diffuse cosmic X-ray background, using the Earth as an occulting device, determined the CXB at hard X-ray energies with high accuracy [16].

• Hard X-ray/soft γ-ray point source catalogues provide important new data on galactic and extragalactic source populations in the 15 keV – 8 MeV band [6], [17], [18].

• Detection of polarization in the high – energy emission of the Crab pulsar and in the prompt emission of gamma - ray burst GRB 041219A, both with SPI and IBIS [19], [20], [21], [22].

• The detection of diffuse hard X-ray emission from the Ophiuchus cluster of galaxies [23].

• Spectral lags, i.e. the time difference in the photon arrival times between high energy (> 50 keV) and low-energy (< 50 keV) events, were observed for a number GRB within the INTEGRAL FOV. Mapping those faint (Epeak < 1 ph cm$^{-2}$s$^{-1}$, 20-200 keV) GRB with a long spectral lag (> 0.7 s) suggests an association of their locations with the plane of the super-galactic structure [24].

• Deeper observation of the bulge and disk of the Galaxy in the light of the 511 keV line show an asymmetry in a sense that higher fluxes are observed at negative galactic longitudes in the plane. The bulge emission remains symmetric [25].

• Deep observations of the galactic ridge emission [26], [27] from the inner radian show that (i) below 50 to 60 keV, accreting white dwarf binaries (CV) dominate the emission which is compatible with the NIR emission at 4.9 μ [26]; (ii) the integrated signal fom point sources dominate the range from 50 keV to ~ 300 keV; (iii) 511 keV line and Ps continuum emission are prominent between 300, and 511 keV; and (iv) a diffuse, hard power-law component emerges at energies above 300 keV. The latter one can be well modeled, together with data from CGRO, by Inverse Compton scattering from relativistic electrons in the ISM, showing that INTEGRAL is a good tracer of GeV electrons and positrons in our Galaxy [28], however, a yet unknown population of sources with hard spectra (Pulsars ? AXP ? CV?) can not entirely be ruled out to contribute to the PL emission in the innter Galaxy.

• The contribution of Compton-thick AGN to the CXB radiation, which has a maximum at around 25 keV is not yet fully understood, as these AGN contribute only a few % to the total observed flux. Recent analysis of the NH distribution in a complete sample of 88 AGN selected in the 20-40 keV shows [29], that at low redshifts (z < 0.015) INTEGRAL sees almost the entire AGN population, from unabsorbed to at least mildly Compton thick objects, while in the total sample, one looses the heavily absorbed 'counterparts' of distant and therefore dim sources with little or no absorption. Taking therefore the low-z bin as the only one being able to provide the 'true' distribution of absorption in type 1 and 2 AGN, the fraction of Compton thick objects is estimated to be >24%.

## 6. Summary and outlook

INTEGRAL, launched on 17 October 2002, has been in its 2-year nominal mission phase of science operations until 01 January 2004. Then the extended science operations phase started and is continuing smoothly. During this time, INTEGRAL, as a mature and efficiently operated mission has provided a wealth of exciting and solid new science. The spacecraft is in good health with sufficient resources. INTEGRAL operates in a unique high-energy band, with SPI as a unique spectrometer for nuclear-line astrophysics and IBIS as the unique imager for high-energy point sources. The observing programmes are oversubscribed and the numbers of published papers are continuously increasing. The INTEGRAL users community plays a very active role and strongly supports the mission.

The extended mission phase of science operations has been approved recently to last until 31 December 2012, subject to a science and performance review scheduled for the Fall of 2010. This 2010 review will also include an indication of a new budget required to operate the mission into 2013 and 2014. It should be noted, that the total costs of INTEGRAL science operations to ESA, per year, amount to about 7.3 M€ per year (2009 economic conditions) which corresponds to about 1% only of the total costs of the mission ("as built") which includes the spacecraft, payload, launcher, ground segment, and ISDC. In return for this 1% input, the science community can benefit from an output of about 15%, which is the science observing time gained, corresponding to about one additional year of science operations. In order to safeguard a continuation of INTEGRAL science operations in the future, it is mandatory that not only the spacecraft and payload continue to operate flawlessly, but that also, as proven before, the high quality of the scientific productivity of this mission is being maintained.

## References

[1] C. Winkler et al., 2003, A&A 411, L1

[2] G. Vedrenne et al., 2003, A&A 411, L63

[3] P. Ubertini et al., 2003, A&A 411, L131

[4] N. Lund et al., 2003, A&A 411, L231

[5] J.M. Mas-Hesse et al., 2003, A&A 411, L261

[6] A.J. Bird et al., 2009, ApJS, in press, arXiv:0910.1704

[7] J.-P. Roques, 2007, INTEGRAL Mission Extended Operations Review, ESA May 2007

[8] R. Walter et al., A&A 411, L427, 2003

[9] A. Lutovinov et al., A&A 444, 821, 2005

[10] A. Bodaghee et al., A&A 467, 585, 2007

[11] F. Lebrun et al., Nature 428, 293, 2004

[12] J. Knödlseder et al., A&A 441, 513, 2005

[13] R. Diehl et al., Nature 439, 45, 2006

[14] L. Kuiper et al., ApJ 645, 556, 2006

[15] V. Sguera et al., ApJ 646, 452, 2006

[16] E. Churazov et al., A&A 467, 529, 2007

[17] L. Bouchet et al., ApJ 679, 1315, 2008

[18] R. Krivonos et al., A&A 475, 775, 2007

[19] A.J. Dean et al., Science 321, 1183, 2008

[20] M. Forot et al., ApJL 688, L29, 2008

[21] D. Götz et al., ApJ 695, L208, 2009

[22] S. McGlynn et al., A&A 466, 895, 2007

[23] D. Eckert et al., A&A 479, 27, 2008

[24] S. Foley et al., A&A 484, 143, 2008

[25] G. Weidenspointner et al., Nature 451, 159, 2008

[26] M. Revnivtsev et al., A&A 489, 1121, 2008

[27] L. Bouchet et al., ApJ 679, 1315, 2008

[28] T. Porter et al., ApJ 682, 400, 2008

[29] A. Malizia et al., MNRAS 399, 944, 2009

[30] E. Churazov et al., MNRAS 357, 1377, 2005

[31] P. Jean et al., A&A 445, 579, 2006